# REVIEW OF METHODS OF POWER-SPECTRUM ANALYSIS AS APPLIED TO SUPER-KAMIOKANDE SOLAR NEUTRINO DATA

By P.A. Sturrock[1]

[1] Center for Space Science and Astrophysics, Varian 302, Stanford University, Stanford, California 94305

To help understand why different published analyses of the Super-Kamiokande solar neutrino data arrive at different conclusions, we have applied six different methods to a standardized problem. The key difference between the various methods rests in the amount of information that each processes. A Lomb-Scargle analysis that uses the mean times of the time bins and ignores experimental error estimates uses the least information. A likelihood analysis that uses the start times, end times, and mean live times, and takes account of the experimental error estimates, makes the greatest use of the available information. We carry out power-spectrum analyses of the Super-Kamiokande 5-day solar neutrino data, using each method in turn, for a standard search band (0 to 50 $yr^{-1}$). For each method, we also carry out a fixed number (10,000) of Monte-Carlo simulations for the purpose of estimating the significance of the leading peak in each power spectrum. We find that, with one exception, the results of these calculations are compatible with those of previously published analyses. (We are unable to replicate Koshio's recent results.) We find that the significance of the peaks at 9.43 $yr^{-1}$ and at 43.72 $yr^{-1}$ increases progressively as one incorporates more information into the analysis procedure.

1. INTRODUCTION

There have been a number of articles presenting the results of power-spectrum analyses of the Super-Kamiokande solar neutrino data. Those published by members of the Super-Kamiokande collaboration claim that there is no evidence for variability [1-3]. On the other hand, investigators outside of the collaboration have claimed to find evidence for variability. Milsztajn [4] claimed to find evidence for variability in an analysis of Super-Kamiokande 10-day data. The Stanford collaboration has presented evidence in favor of variability from analyses of both the 10-day data [5 - 7] and the 5-day data [8 - 10]. One difference is that early articles worked with the first dataset, which was organized in 10-day bins, whereas later analyses used the second dataset, which was organized in 5-day bins. However, the methodologies have not been uniform (including differences in significance estimation, search bands, etc.), so it is important to understand the extent to which the differences in the claimed conclusions may be attributed to methodological differences.

For this reason, we have carried out a sequence of power-spectrum analyses of the Super-Kamiokande 5-day dataset, using in turn the five methods that have so far been used, and one more that one might have expected to have been used. We standardize other factors by (a) concentrating on the 5-day dataset, (b) adopting a standard search band (0 to 50 $yr^{-1}$), and (c) using Monte-Carlo simulations for significance estimates. The 5-day dataset [2] lists the start times, end times, mean live times, flux estimates, lower error estimates, and upper error estimates for each of 358 bins, beginning on May 31, 1996, and ending on July 15, 2001. We order our power-spectrum analyses in a sequence that uses progressively more experimental information.

In Section 2, we give the results of a simple Lomb-Scargle analysis [11, 12] of the 5-day dataset, using only the mean times and the flux estimates, as in the Milsztajn [4] analysis of the Super-Kamiokande 10-day data, standardizing the frequency interval to be 0 to 50 $yr^{-1}$. We also present the results of Monte-Carlo simulations (similar to analyses carried out by Nakahata [1], Yoo [2], and Koshio [3]). In Section 3, we repeat the analysis of Section 2, but now incorporate the mean live times rather than the mean times, as in the analyses of Nakahata [1] and Yoo [2]. We again present the results of Monte-Carlo simulations. We repeat the above calculations in Section 4, but follow Scargle [13] in extending the basic Lomb-Scargle method to take account of the experimental error estimates. This results in significant changes in the power spectrum and in the implications of the Monte-Carlo simulations. The biggest peak is now that at 9.43 yr$^{-1}$ with power 9.56.

The next step, in Section 5, is to analyze the data by a likelihood method introduced by Sturrock, Walther, and Wheatland [14], which we refer to as the "SWW likelihood method." In this section, we take account of the start times, end times, flux estimates, and error estimates. The leading peak remains that at 9.43 yr$^{-1}$ and we find from examination of the power spectrum and from Monte-Carlo simulations that the significance of this peak is increased. In Section 6, we use the SWW likelihood method in such a way as to take account of the mean live times as well as the start times, end times, flux estimates, and error estimates. We find from the power spectrum and from Monte-Carlo simulations that the significance of the peak at 9.43 yr$^{-1}$ is further increased.

In Section 7, we analyze the Super-Kamiokande 10-day dataset, using the method used in Section 6. We compare the results with those obtained from the 5-day dataset, and point out that the main difference is due to aliasing. In Section 8, we discuss a procedure, recently used by Koshio [3], that involves a least-squares fit to the data, adjusting the complex amplitude (as in the SWW likelihood method), but also allowing for a variable offset. We find that our results are inconsistent with those published by Koshio, but they are consistent with the results of earlier sections of this article. Final comments are given in Section 9. In an appendix, we comment on the concept of "independent frequencies" used by Milsztajn [4] and Yoo et al. [2] in estimating the significance of peaks in the power spectrum.

## 2. LOMB- SCARGLE ANALYSIS, USING THE MEAN TIMES

The Super-Kamiokande 5-day data are organized in bins which we enumerate by r = 1,…,R, where R = 358. For each bin we are given the start time $t_{s,r}$, the end time, $t_{e,r}$, the "weighted mean live time" $t_{ml,r}$, the flux estimate $g_r$, the lower error estimate $s_{l,r}$, and the upper error estimate $s_{u,r}$. We find that the two error estimates have a close relationship: their ratio has a range of only a few percent. For this reason, we here work with a single error estimate formed from their mean:

$$s_r = \tfrac{1}{2}(s_{l,r} + s_{u,r}). \tag{2.1}$$

We now normalize the flux estimates

$$x_r = \frac{g_r}{mean(g_s)} - 1, \tag{2.2}$$

and also the error estimates

$$\sigma_r = \frac{s_r}{mean(g_s)}, \tag{2.3}$$

but the experimental error estimates are not used in the basic Lomb-Scargle calculations of this section.

Following Lomb [11] and Scargle [12] (see also Press et al. [15]), we form a power spectrum from

$$S(\nu) = \frac{1}{2\sigma_0^2}\left\{\frac{\left[\sum_r x_r \cos(2\pi\nu(t_r - \tau))\right]^2}{\left[\sum_r \cos^2(2\pi\nu(t_r - \tau))\right]} + \frac{\left[\sum_r x_r \sin(2\pi\nu(t_r - \tau))\right]^2}{\left[\sum_r \sin^2(2\pi\nu(t_r - \tau))\right]}\right\}, \tag{2.4}$$

where

$$\sigma_0 = std(x_r), \tag{2.5}$$

and $\tau$ is defined by the relation

$$\tan(4\pi\nu\tau) = \frac{\sum_r \sin(4\pi\nu t_r)}{\sum_r \cos(4\pi\nu t_r)}. \tag{2.6}$$

In order to use the Lomb-Scargle procedure, it is necessary to assign a definite time $t_r$ to each bin. In this section, we adopt the mean of the start and end time, as in the early work of Milszajn [4]. This yields the power spectrum shown in Figure 1. The top ten peaks are listed in Table 1. Of the five leading peaks, those at frequencies 9.43 yr$^{-1}$, 39.27 yr$^{-1}$, and 43.73 yr$^{-1}$ recur in later analyses.

Here and in later sections, we assess the significance of the leading peak by Monte-Carlo simulations. We generate a large number of simulated datasets by the algorithm

$$x_{MC,r} = \sigma_r randn, \qquad (2.7)$$

where randn is the operation of producing random numbers with a normal distribution and variance unity. (Yoo et al. [2] use (effectively) $\sigma_0$ in their Monte Carlo calculations, but we use (2.7) consistently throughout this article.) For each fictitious dataset, we compute the power spectrum over the range 0 to 50 yr$^{-1}$, and note the power SM of the biggest peak. We then examine the distribution of the maximum-power values.

We present the results of these simulations in two ways. Figure 2 presents the reverse cumulative distribution function that shows, as ordinate, the logarithm of the fraction of the simulations with power exceeding the value shown in the abscissa. We see that 49% of the simulations have power equal to or exceeding that of the strongest peak (S = 6.79) in the actual power spectrum. From this way of looking at the data, one would conclude that there is no evidence for a periodic modulation of the neutrino flux. In order to facilitate the comparison of our results with those of the Super-Kamiokande collaboration, we also present the results of the Monte Carlo simulations in histogram form. Figure 3 shows the distribution of values of SM from the simulations, and indicates the value of SM for the actual data.

It is important to note that this test assumes that, a priori, all frequencies in the chosen search band (here 0 to 50 yr$^{-1}$) are equally likely. Hence we are ignoring all available information concerning variability in solar structure and dynamics. We comment on this point further in Section 9.

## 3. LOMB-SCARGLE ANALYSIS, USING THE MEAN LIVE TIMES

We now repeat the analysis of Section 2, referring measurements to the mean live times, rather than the mean times. The power spectrum, computed again by the basic Lomb-Scargle method, is shown in Figure 4, and the top ten peaks are listed in Table 2. Figure 5 presents the reverse cumulative distribution, and figure 6 presents the same data in histogram form. We see that 32% of the simulations have power equal to or exceeding that of the strongest peak (power 7.29 at frequency 43.73 yr$^{-1}$) in the actual power spectrum. From this way of looking at the data, one would conclude that there is no evidence for a periodic modulation of the neutrino flux.

4. MODIFIED LOMB-SCARGLE ANALYSIS, USING THE MEAN LIVE TIMES

We now repeat the calculations of the previous section, replacing the basic Lomb-Scargle procedure by a modified procedure that takes account of the experimental error estimates. Following Scargle [13], we introduce a weighting term given by

$$w_r = \frac{1/\sigma_r^2}{mean(1/\sigma_s^2)} \;. \qquad (4.1)$$

We then replace (2.4) by

$$S(\nu) = \frac{1}{2\sigma_0^2} \left\{ \frac{\left[\sum_r w_r x_r \cos(2\pi\nu(t_r - \tau))\right]^2}{\left[\sum_r w_r \cos^2(2\pi\nu(t_r - \tau))\right]} + \frac{\left[\sum_r w_r x_r \sin(2\pi\nu(t_r - \tau))\right]^2}{\left[\sum_r w_r \sin^2(2\pi\nu(t_r - \tau))\right]} \right\}, \qquad (4.2)$$

where $\sigma_0$ and $\tau$ are now defined by

$$\sigma_0 = std(w_r x_r) \qquad (4.3)$$

and $$\tan(4\pi\nu\tau) = \frac{\sum_r w_r \sin(4\pi\nu t_r)}{\sum_r w_r \cos(4\pi\nu t_r)}. \qquad (4.4)$$

When we apply this procedure to the Super-Kamiokande 5-day dataset (now taking account of the mean live times, the flux estimates, and the error estimates), we obtain the power spectrum shown in Figure 7. The top ten peaks are listed in Table 3. We see that the three most significant peaks in this power spectrum are those at frequencies 9.43 yr$^{-1}$, 43.72 yr$^{-1}$, and 39.28 yr$^{-1}$, with powers 9.56, 7.91, and 6.18, respectively. The results of Monte Carlos simulations are shown (in reverse cumulative distribution form) in Figure 8, and (in histogram form) in Figure 9. We find that less than 5% of the simulations (477 out of 10,000) have power equal to or larger than the actual maximum power in the range 0 to 50 yr$^{-1}$, i.e. 9.56, which is found at frequency 9.43 yr$^{-1}$.

## 5. SWW LIKELIHOOD ANALYSIS, USING THE START TIMES AND END TIMES

We now carry out power spectrum analysis using the SWW likelihood procedure [14]. Using the notation of Section 2, the log-likelihood that the data may be fit to a model that gives $X_r$ as the expected values of $x_r$ is given by

$$L = -\tfrac{1}{2}\sum_{r=1}^{R}(x_r - X_r)^2 / \sigma_r^2. \qquad (5.1)$$

We estimate the power spectrum from the increase in the log-likelihood over the value expected for no modulation, corresponding to $X_r = 0$:

$$S = \tfrac{1}{2}\sum_{r=1}^{R}\frac{x_r^2}{\sigma_r^2} - \tfrac{1}{2}\sum_{r=1}^{R}\frac{(x_r - X_r)^2}{\sigma_r^2}. \qquad (5.2)$$

In this section, we assume that the data-acquisition process is uniform over the duration of each bin, and we examine the possibility that the flux varies sinusoidally with frequency $\nu$. Then the expected normalized flux estimates will be given by

$$X_r = \frac{1}{D_r}\int_{t_{sr}}^{t_{er}} dt \left(A e^{i2\pi\nu t} + A^* e^{-i2\pi\nu t}\right), \qquad (5.3)$$

where

$$D_r = t_{e,r} - t_{s,r} \qquad (5.4)$$

and, for each frequency, the complex amplitude A is adjusted to maximize the likelihood.

The resulting power spectrum is shown in Figure 10, and the top ten peaks are listed in Table 4. The results of Monte Carlo simulations are shown in Figure 11 (in reverse cumulative distribution form), and in Figure 12 (in histogram form). We find that less than 1% of the simulations (90 out of 10,000) have power equal to or larger than the actual maximum power in the range 0 to 50 yr$^{-1}$, i.e. 11.51, which is found at frequency 9.43 yr$^{-1}$.

## 6. SWW LIKELIHOOD ANALYSIS, USING THE START TIMES, END TIMES, AND MEAN LIVE TIMES

We now modify the SWW likelihood procedure in such a way as to allow us to take account of the mean live times, as well as the start times and end times. We now replace equation (5.3) by

$$X_r = \frac{1}{D_r} \int_{t_{sr}}^{t_{er}} dt W_r(t) \left( A e^{i2\pi\nu t} + A^* e^{-i2\pi\nu t} \right), \tag{6.1}$$

where the weighting function $W_r(t)$ is chosen so that the mean value is unity, but

$$\frac{1}{D_r} \int_{t_{sr}}^{t_{er}} dt W_r(t) t = t_{ml}. \tag{6.2}$$

We may meet these requirements with the following simple "double-boxcar" model:

$$W_r(t) = W_{l,r} \equiv \frac{t_{e,r} - t_{ml,r}}{t_{ml,r} - t_{s,r}} \text{ for } t_{s,r} < t < t_{ml,r}$$

$$W_r(t) = W_{u,r} \equiv \frac{t_{ml,r} - t_{s,r}}{t_{e,r} - t_{ml,r}} \text{ for } t_{ml,r} < t < t_{e,r} \tag{6.3}$$

We have used this modification of the SWW likelihood method to compute the power spectrum of the Super-Kamiokande 5-day data. The result is shown in Figure 13, and the top ten peaks are listed in Table 5. The results of Monte Carlo simulations are shown in Figure 14 (in reverse cumulative distribution form), and in Figure 15 (in histogram form). We again find that less than 1% of the simulations (now 74 out of 10,000) have power equal to or larger than the actual maximum power in the range 0 to 50 yr$^{-1}$, i.e. 11.67, which is found at frequency 9.43 yr$^{-1}$.

## 7. SWW LIKELIHOOD ANALYSIS OF THE SUPER-KAMIOKANDE 10-DAY DATA, USING THE START TIMES, END TIMES, AND MEAN LIVE TIMES

In order to understand the relationship of power spectra formed from the Super-Kamiokande 10-day and 5-day datasets, it is useful to apply the analysis of Section 6 to the 10-day dataset. The resulting power spectrum is shown in Figure 16, and the top ten peaks are listed in Table 6.

We see that the principal peak in the power spectrum is found at frequency 26.57 yr$^{-1}$, with power 11.26. The second peak is at 9.42 yr$^{-1}$ with power 7.29. As we have pointed out elsewhere [9], the difference between the 5-day and 10-day power spectra is due to aliasing. If the power spectrum of the bin-times contains a peak at frequency $\nu_T$, and if the data contains modulation at frequency $\nu_M$, then the power spectrum will also exhibit peaks at $|\nu_T - \nu_M|$ and at $\nu_T + \nu_M$. If the peak at $\nu_T$ is particularly strong (as it is for the Super-Kamiokande datasets), the power spectrum may also exhibit peaks at $|2\nu_T - \nu_M|$ and $2\nu_T + \nu_M$, etc. For the 10-day dataset, $\nu_T = \nu_{T10} \approx 36$ yr$^{-1}$. Since $9.42 + 26.57 = 35.99$, we may conclude that the peaks at 9.42 yr$^{-1}$ and 26.57 yr$^{-1}$ are related, one being an alias of the other. When only the 10-day dataset was available, it seemed reasonable to guess that the primary peak was that at 26.57 yr$^{-1}$ since that was the stronger of the two and could be interpreted as the second harmonic of the synodic solar rotation frequency. However, analysis of the 5-day dataset has made it clear that the reverse is the case: the primary peak is that at 9.42 or 9.43 yr$^{-1}$. This explains why the peak at 26.57 yr$^{-1}$ appears in the power spectrum formed from the 10-day dataset, but not in that formed from the 5-day dataset.

We may make a more objective assessment of the role of aliasing in the power spectrum formed from the 10-day dataset by using the "joint power statistic," that provides a convenient procedure for examining the correlation of two power spectra [17]. If we form the geometric mean of the powers,

$$X = (S_1 S_2)^{1/2} , \qquad (7.1)$$

the joint power statistic (of second order) is given by

$$J = -\ln(2 X K_1(2X)) \qquad (7.2)$$

where $K_1$ is the Bessel function of the second kind. This function has the following useful property: if $S_1$ and $S_2$ are distributed exponentially, then J also is distributed exponentially. Hence a display of J may be interpreted in the same way as a display of a power spectrum.

Figure 17 shows the joint power statistic formed from $S(\nu)$ and $S(\nu_T - \nu)$, over the frequency range 0 to 18 yr$^{-1}$. It is clear from this display that the peaks at 9.42 yr$^{-1}$ and at 26.57 yr$^{-1}$ are indeed an alias pair. Similar analysis of the 5-day dataset shows that aliasing again plays a role, but is much less important since the timing frequency is much higher (about 72 yr$^{-1}$).

It is interesting to compare the 5-day and 10-day power spectra in more detail. If the solar-neutrino flux is in fact modulated sinusoidally with a well-defined frequency, we would expect to find peaks at that frequency in both power spectra. However, since there is more information in the 5-day dataset, it is reasonable to expect that the peak in that power spectrum will be more pronounced than that in the 10-day power spectrum. Figure 18 is a plot of the peaks in the two power spectra, where the abscissa values are the powers in the 10-day dataset, and the ordinate values are the powers in the 5-day dataset. We see that there are four peaks that are "outside the pack." One of these has high power in the 10-day spectrum, but only weak power in the 5-day spectrum. This is the peak at 25.57 yr$^{-1}$ which we have just discussed. The other three peaks (at 9.43 yr$^{-1}$, 39.28 yr$^{-1}$, and 43.72 yr$^{-1}$) appear in both the 10-day and 5-day spectra, and they are indeed stronger in the latter than in the former, consistent with the possibility that they represent real modulation.

## 8. LIKELIHOOD ANALYSIS WITH FLOATING OFFSET

Koshio [3] has recently published an analysis of Super-Kamiokande 5-day data using a method that is the same as the SWW likelihood method if one takes account of the start times, end times, and error estimates (but not the mean live times), and if one adjusts not only the complex amplitude for each frequency, but also the offset. Then Equation (5.2) is replaced by

$$S = \tfrac{1}{2}\sum_{r=1}^{R}\frac{g_r^2}{s_r^2} - \tfrac{1}{2}\sum_{r=1}^{R}\frac{(g_r - G_r)^2}{s_r^2} \tag{8.1}$$

where

$$G_r = \frac{1}{D_r}\int_{t_{sr}}^{t_e} dt\left(C + Ae^{i2\pi\nu t} + A^* e^{-i2\pi\nu t}\right), \tag{8.2}$$

and we adjust both C and A, for each frequency, to maximize S. We refer to this as the "floating offset" procedure. We show in Figure 19 the power spectrum obtained from this procedure. It is quite consistent with the power spectrum computed by Koshio [3].

We have carried out 10,000 Monte Carlo simulations of this calculation, with results shown in reverse-cumulative-distribution form in Figure 20, and in histogram form in Figure 21. We find that only 251 out of 10,000 simulations have power as large as or larger than the actual maximum power (11.24 at frequency 9.43 yr$^1$). This fraction (2.5%) is much smaller than the value (20.94%) given by Koshio on the basis of his Monte Carlo simulations. Unfortunately, there is insufficient information in Koshio's article to enable one to understand the source of this discrepancy.

However, it is important to try to understand why this calculation gives results that are less significant than those obtained by the SWW likelihood method used in Section 5, that also uses the start times, end times, and error estimates, but not the mean live times. One might expect that one would obtain a "better" estimate of the power spectrum by seeking a more flexible fit. However, on examining the calculation is more detail, we find cause for concern. It is clear that the equations are ill conditioned at zero-frequency, since the maximum-likelihood procedure can determine $C + 2|A|$, but not C or A separately. It appears that a similar difficulty plays a role at non-zero frequencies also. We see in Figure 22 that the amplitude estimates become quite erratic for frequencies higher than 30 yr$^{-1}$. The same is true of the offset estimates. For this reason, it seems safer not to incorporate a floating offset in power spectrum calculations.

## 9. DISCUSSION

Although we have presented the analyses of Sections 2 and 3 in their conventional forms, and introduced the likelihood procedure of Sections 4, 5 and 6 as something different, we may in fact regard all the analyses presented in this article as special cases of the likelihood procedure. The relationship is shown schematically in Figure 23. Panels (a) and (b) show the "single boxcar" and "double boxcar" weighting functions corresponding to the uniform weighting in Equation (5.3) and the non-uniform weighting in Equation (6.3). If one calculates the power spectrum from the SWW likelihood procedure, adopting the standard deviation of the flux estimates as the error term and using a delta-function form of the time weighting function, as in panels (c) and (d), one retrieves the power spectra computed by the Lomb-Scargle procedure in Sections 2 and 3, respectively. The calculation of Section 4 is equivalent to using the time weighting function shown in panel (d) and the actual error estimates.

We now compare the results of these analyses with those of previous publications. Milsztajn [4] used the basic Lomb-Scargle method to analyze the 10-day dataset, assigning flux measurements to the mean times. Hence his method was that of Section 2, but it was applied to the dataset analyzed in Section 7. Milsztajn's power spectrum is in fact very close to that obtained in Section 6 (see Figure 16). The two principal peaks are found at frequencies 26.57 $yr^{-1}$ and 9.42 $yr^{-1}$. In his article, Milsztajn states "…the sampling, though quite regular, is sufficiently variable that no aliasing is observed…" but is incorrect. We saw in Section 6 that the two principal peaks comprise an alias pair, related by the timing frequency 35.99 $yr^{-1}$.

In our earlier analysis of the 10-day dataset [7], we obtained a power spectrum very similar to that found by Milsztajn from an SWW likelihood analysis that took account of the start times, end times, and error estimates. This produced a power spectrum (similar but not identical to that shown as Figure 16) that has six peaks with S > 5 over the frequency range 0 to 40 $yr^{-1}$, However, when we determined the amplitude and phase of the principal modulation at 26.57 $yr^{-1}$ and subtracted this modulation from the dataset, we found that the power spectrum of the resulting time series has no peak with S > 5. This fact also is indicative of aliasing due to the near-regularity of the binning.

Nakahata [1] also carried out a Lomb-Scargle analysis of the 10-day dataset, assigning measurements to the mean times, and obtained a power spectrum close to those found previously [4, 7]. However, Nakahata had access to the mean live time measurements, and therefore repeated the Lomb-Scargle analysis, assigning flux measurements to the mean live times rather than to the mean times. This analysis yields a peak at frequency 26.55 $yr^{-1}$ with power 7.51, and a peak at frequency 9.42 $yr^{-1}$ with power 6.67. Nakahata interprets the second peak as "a natural peak in the random distribution," whereas, as we have seen in Section 7, it is in fact an alias of the first peak.

Yoo et al. [1] were the first to have access to and to analyze the 5-day dataset. Their analysis is that reproduced in Section 3, leading to the power spectrum shown in Figure 4. Yoo et al. commented on our analysis of the 10-day dataset [7, 8] and asserted that the difference in the resulting power spectra was due to the fact that our analysis used the mean times rather than the mean live times, but this statement was incorrect, since our analysis used the start times and end times, and made no reference to the mean times. Yoo et al. noted that the peak at 26.55 $yr^{-1}$, which was evident in the 10-day power spectrum, was no longer evident in the 5-day power spectrum, and concluded that this "provides additional confirmation that the [peak at 26.55 $yr^{-1}$] in the 10-day long sample is a statistical artifact." However, as we have seen, the peak disappeared because it is an alias in the 10-day power spectrum but not in the 5-day power spectrum.

In our analysis of the 5-day dataset [9] that uses the procedure summarized in Section 6, we point out that the two strongest peaks (at 9.43 $yr^{-1}$ and 43.72 $yr^{-1}$) may both be due to an internal r-mode oscillation [18, 19, 20] with indices $l = 2$, $m = 2$.

Koshio [3] has recently published an analysis that we attempted to reproduce in Section 8. However, as we point out in that section, our analysis yields evidence for a significant modulation, whereas Koshio's analysis does not.

We now review as a sequence the analyses presented in Sections 2 through 6. The principal peak in the power spectrum formed in Section 6 (which takes account of more information than the earlier analyses of Sections 2 through 5) is that at 9.43 $yr^{-1}$. On examining Tables 1 through 5, we find that the power of this peak increases progressively as we analyze the dataset by procedures that use progressively more information. This is what one would expect if this peak were due to a real modulation. However, it might be a trend that occurs for all peaks (real or not). It is therefore interesting to compare the trend for the frequencies of the top ten peaks in Table 1. This comparison is shown in Figure 24. We find that, of these ten frequencies, only two show a monotonic increase in power: those at 9.43 yr$^{-1}$ and 43.72 yr$^{-1}$. These are also the two most notable peaks in the display of Figure 18.

We now comment further on a point raised in Section 2:- the Monte Carlo analysis of Nakahata [1], Yoo [2], and Koshio [3] implicitly assumes that, a priori, all frequencies in the chosen search band (here 0 to 50 yr$^{-1}$) are equally likely. This may be appropriate if one has no idea what frequencies are likely to turn up. However, if one were considering the specific possibility that the solar neutrino flux may exhibit a periodic modulation due (for instance) to solar rotation, then the appropriate search band would be determined by the Sun's internal quasi-equatorial synodic rotation (approximately 12 to 14 yr$^{-1}$) [16], and possibly multiples of this range. If one were looking for other types of modulation, such as r-modes, one would need to determine search bands appropriate for those modulations. This is the approach that we have used in previous articles (for instance, [7] and [9]).

There is another basic point that is worth noting. Power spectrum analysis of solar neutrino data may detect an oscillation in the neutrino flux if the flux is modulated by a stable, high-Q, oscillation. However, the flux may be variable, but the variability may not meet these criteria, in which case the variability may well escape detection by power-spectrum analysis. For instance, power-spectrum analysis is not well suited to the detection of a stochastic variation, and it may fail to detect an oscillation that drifts in frequency and/or jumps in phase. Hence even if a power spectrum analysis were to fail to reveal a peak in a wide frequency range (which it does not), this in itself would not comprise evidence that the flux does not vary.

In order to pursue further the question of variability of the solar neutrino flux, it would be most helpful if the SNO collaboration were to make their data publicly available. It would be especially helpful if the Super-Kamiokande and SNO collaborations would provide their data in identical 1-day bins.

This work was supported by NSF grant AST-0097128. I am indebted to David Caldwell, Jeff Scargle, Guenther Walther, and Mike Wheatland, for helpful comments and advice.

## APPENDIX. INDEPENDENT FREQUENCIES

Milsztajn [4] and Yoo et al. [2] have made significance estimates by using the "false-alarm" formula [12, 15]

$$P = 1 - \left(1 - e^{-S}\right)^M \quad (A.1)$$

where M is the number of "independent frequencies" in the search band. In terms of the "timing frequency" $\nu_T$, which is approximately $1/\delta t$ where $\delta t$ is the mean bin duration, both authors adopt as their search band the range $0 - \nu_T$. The timing frequency is twice the Nyquist frequency, given by

$$\nu_{Nyq} = \frac{1}{2\,\delta t} \,. \quad (A.2)$$

For regular sampling, the number of independent frequencies up to the Nyquist limit is the number of samples N. Hence, as Milsztajn and Yoo et al. point out, the number of independent frequencies M is 2N for the search band $0 - \nu_T$. Milsztajn [4] applies this formula to analysis of the 10-day dataset, leading to $M = 368$, and Yoo et al. [4] apply this formula to analysis of the 5-day dataset, leading to $M = 716$.

Since the sampling is not completely regular, 2N is probably not a precise estimate of M. The purpose of this appendix is to point out that one may derive an empirical estimate of M (for any frequency range) from Monte Carlo simulations, since the reverse cumulative distribution provides an estimate of P as a function of S. We show in Figure 25 a plot of the reverse cumulative distribution of the maximum power over the range $0 - 70\,yr^{-1}$, derived from the 5-day dataset by the method of Section 3. The figure also shows the least-squares fit derived from equation (A.1) by adjusting M. The best fit is found for M = 918, which may be compared with the value 716 assumed by Yoo et al. [2].

For this application, the improved estimate of M makes little difference to the conclusions. Considering the peak at 43.73 yr$^{-1}$ with S = 7.29, it changes P from 38.7% to 46.6%. However, in other analyses of other datasets, the difference could be significant. Note also that this procedure (that is applicable even if the sampling is highly irregular) offers a convenient way of extending estimates made from Monte-Carlo simulations. This could be useful if a peak is highly significant, so that a direct test would need an inordinately large number of simulations.

TABLE 1

Top ten peaks in the basic Lomb-Scargle power spectrum,

using the mean times of bins

| Order | Frequency | Power |
|-------|-----------|-------|
| 1 | 43.72 | 6.79 |
| 2 | 34.02 | 6.19 |
| 3 | 39.28 | 6.03 |
| 4 | 31.23 | 5.95 |
| 5 | 9.43 | 5.90 |
| 6 | 12.31 | 5.67 |
| 7 | 39.54 | 5.65 |
| 8 | 48.16 | 4.75 |
| 9 | 0.36 | 4.64 |
| 10 | 15.73 | 4.35 |

TABLE 2

Top ten peaks in the basic Lomb-Scargle power spectrum,

using the mean live times of bins

| Order | Frequency | Power |
|-------|-----------|-------|
| 1 | 43.73 | 7.29 |
| 2 | 34.01 | 6.65 |
| 3 | 9.43 | 6.18 |
| 4 | 39.27 | 5.82 |
| 5 | 12.31 | 5.48 |
| 6 | 39.54 | 5.34 |
| 7 | 48.15 | 5.18 |
| 8 | 31.23 | 4.67 |
| 9 | 0.36 | 4.64 |
| 10 | 15.73 | 4.06 |

TABLE 3

Top Ten Peaks in the Modified Lomb-Scargle Power Spectrum

| Order | Frequency | Power |
|---|---|---|
| 1 | 9.43 | 9.56 |
| 2 | 43.72 | 7.91 |
| 3 | 39.28 | 6.18 |
| 4 | 33.99 | 5.42 |
| 5 | 45.85 | 5.42 |
| 6 | 12.31 | 4.86 |
| 7 | 8.30 | 4.38 |
| 8 | 0.34 | 4.26 |
| 9 | 31.25 | 4.23 |
| 10 | 35.04 | 4.15 |

TABLE 4

Top Ten Peaks in the SWW Likelihood Power Spectrum,

using start and end times

| Order | Frequency | Power |
|---|---|---|
| 1 | 9.43 | 11.51 |
| 2 | 43.72 | 9.83 |
| 3 | 39.28 | 8.91 |
| 4 | 48.43 | 6.57 |
| 5 | 12.31 | 6.21 |
| 6 | 31.24 | 6.20 |
| 7 | 45.86 | 6.20 |
| 8 | 34.00 | 5.83 |
| 9 | 48.16 | 5.78 |
| 10 | 39.55 | 5.49 |

TABLE 5

Top Ten Peaks in the SWW Likelihood Power Spectrum,

using start times, end times, and mean live times

| Order | Frequency | Power |
|---|---|---|
| 1 | 9.43 | 11.67 |
| 2 | 43.72 | 9.87 |
| 3 | 39.28 | 8.18 |
| 4 | 48.43 | 6.72 |
| 5 | 33.99 | 6.58 |
| 6 | 48.16 | 6.09 |
| 7 | 12.31 | 6.05 |
| 8 | 48.69 | 5.84 |
| 9 | 37.12 | 5.65 |
| 10 | 8.30 | 5.32 |

TABLE 6

Top Ten Peaks in the SWW Likelihood Power Spectrum computed from the 10-day dataset,

using start times, end times, and mean live times

| Order | Frequency | Power |
|---|---|---|
| 1 | 26.57 | 11.26 |
| 2 | 9.42 | 7.29 |
| 3 | 43.73 | 6.55 |
| 4 | 27.02 | 6.15 |
| 5 | 23.63 | 5.38 |
| 6 | 12.36 | 5.30 |
| 7 | 39.31 | 5.28 |
| 8 | 39.59 | 5.26 |
| 9 | 8.31 | 5.05 |
| 10 | 11.56 | 4.91 |

TABLE 7

Top Ten Peaks in the SWW Likelihood Power Spectrum,
with floating offset, using start times and end times

| Order | Frequency | Power |
|---|---|---|
| 1 | 9.43 | 11.24 |
| 2 | 43.72 | 9.44 |
| 3 | 39.28 | 8.64 |
| 4 | 48.43 | 6.38 |
| 5 | 45.86 | 6.10 |
| 6 | 31.24 | 6.03 |
| 7 | 12.31 | 6.01 |
| 8 | 48.16 | 5.69 |
| 9 | 33.99 | 5.63 |
| 10 | 39.55 | 5.32 |

FIGURES

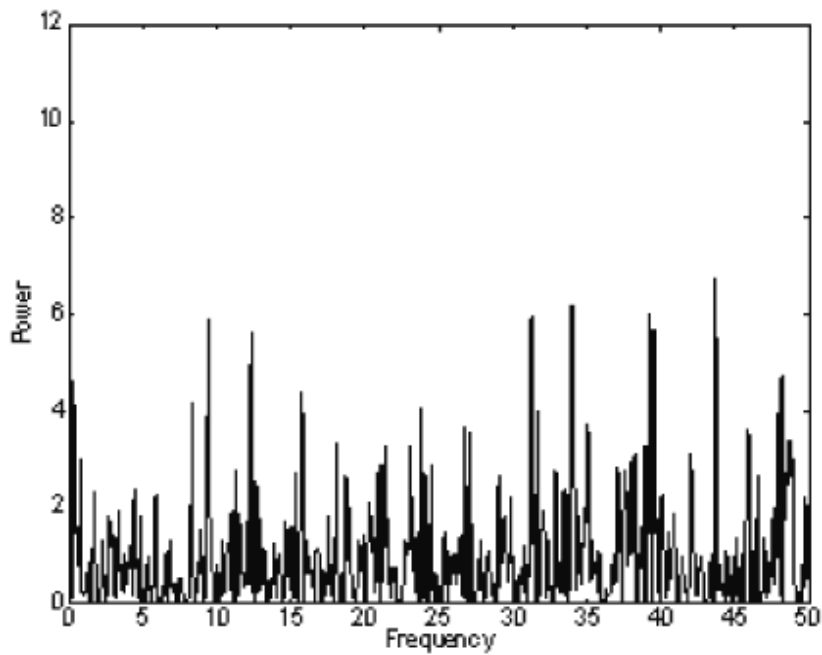

Figure 1. Power spectrum of 5-day Super-Kamiokande data formed by the basic Lomb-Scargle method, using the mean times.

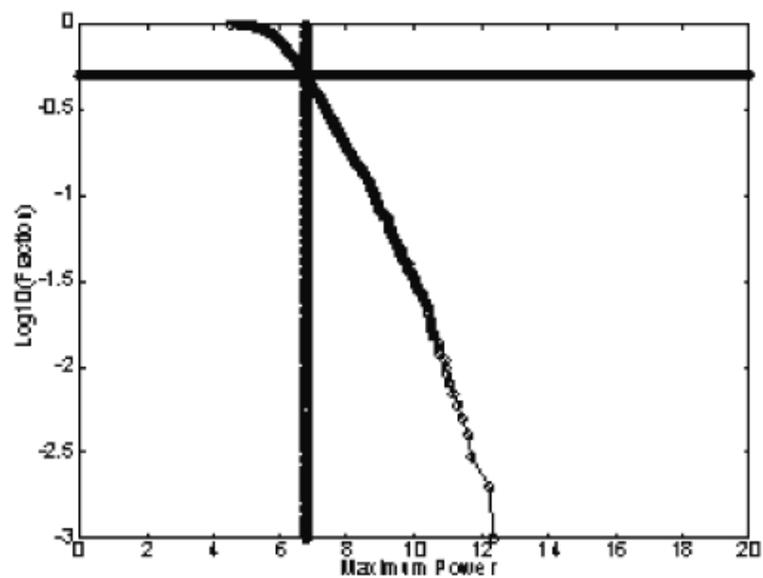

Figure 2. Reverse cumulative distribution function for maximum power, computed by the Lomb-Scargle procedure, using the mean times, over the frequency band 0 to 50 yr$^{-1}$, for 1,000 Monte Carlo simulations of the Super-Kamiokande 5-day data. 485 out of 1,000 simulations have power larger than the actual maximum power (6.79 at frequency 43.72 yr$^{-1}$).

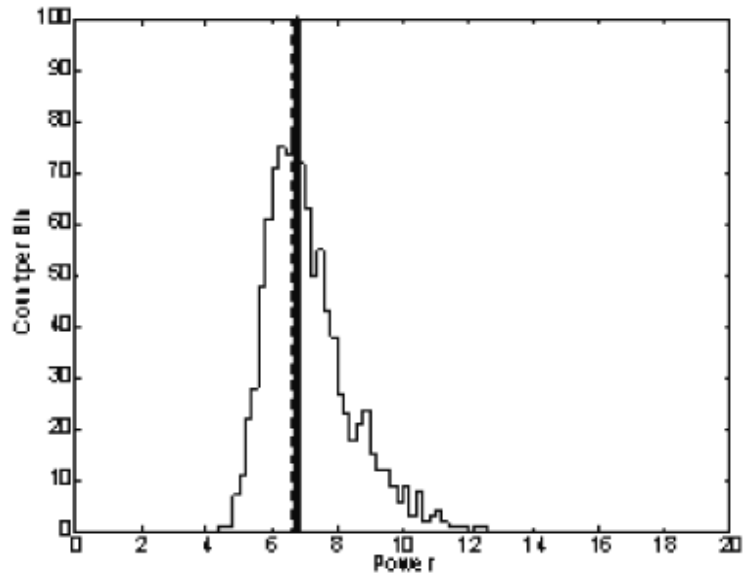

Figure 3. Histogram display of the maximum power, computed by the Lomb-Scargle procedure using the mean times, over the frequency band 0 to 50 yr$^{-1}$, for 1,000 Monte Carlo simulations of the Super-Kamiokande 5-day data. 485 out of 1,000 simulations have power larger than the actual maximum power (6.79 at frequency 43.72 yr$^{-1}$).

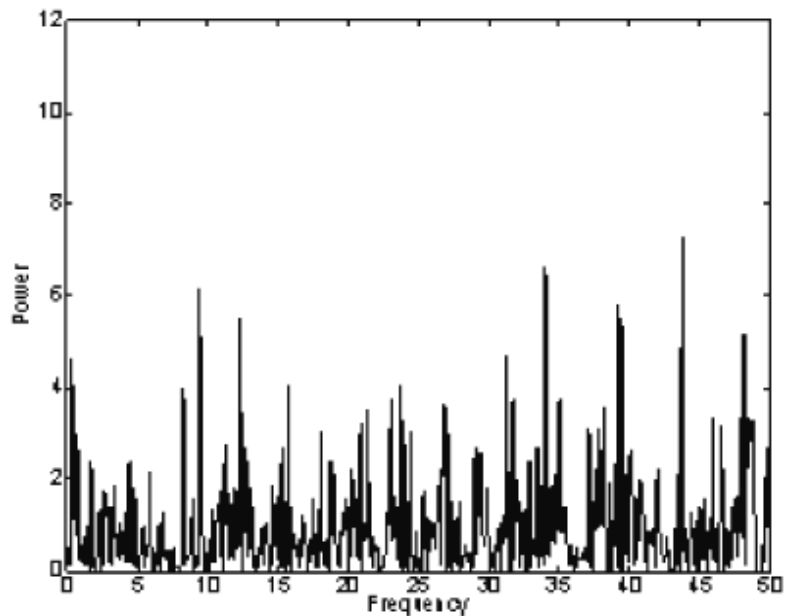

Figure 4. Power spectrum of 5-day Super-Kamiokande data formed by the basic Lomb-Scargle method, using the mean live times.

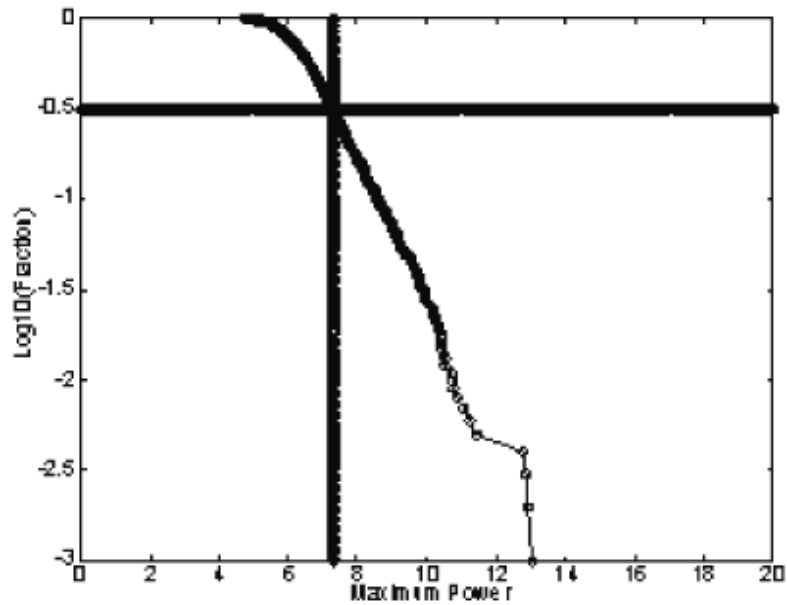

Figure 5. Reverse cumulative distribution function for maximum power, computed by the Lomb-Scargle procedure, using the mean live times, over the frequency band 0 to 50 yr$^{-1}$, for 1,000 Monte Carlo simulations of the Super-Kamiokande 5-day data. 315 out of 1,000 simulations have power larger than the actual maximum power (7.29 at frequency 43.73 yr$^{1}$).

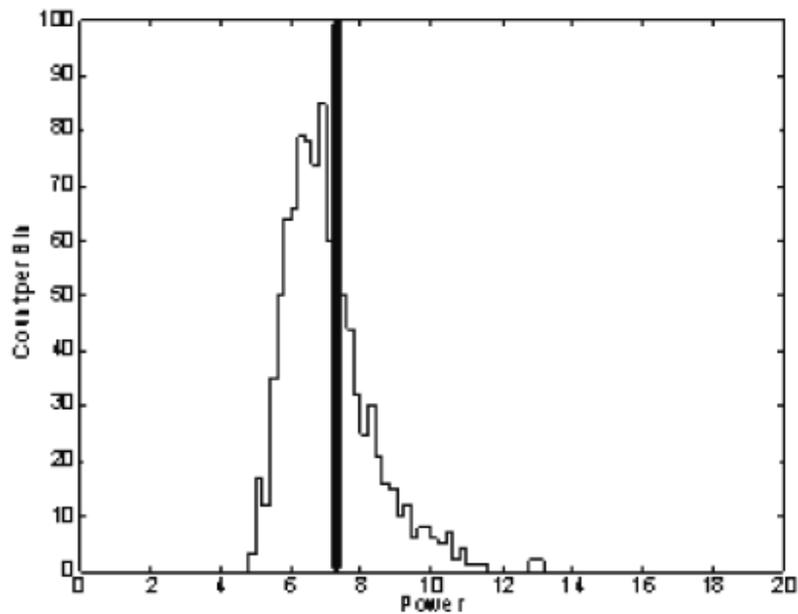

Figure 6. Histogram display of the maximum power, computed by the Lomb-Scargle procedure using the mean live times, over the frequency band 0 to 50 yr$^{-1}$, for 1,000 Monte Carlo simulations of the Super-Kamiokande 5-day data. 315 out of 1,000 simulations have power larger than the actual maximum power (7.29 at frequency 43.73 yr$^{1}$).

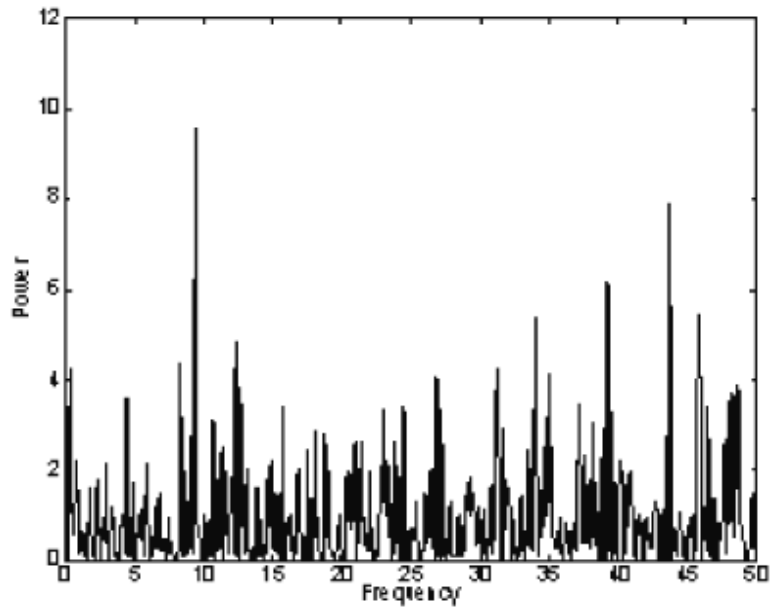

Figure 7. Power spectrum of 5-day Super-Kamiokande data, using the mean live times, formed by the modified Lomb-Scargle method.

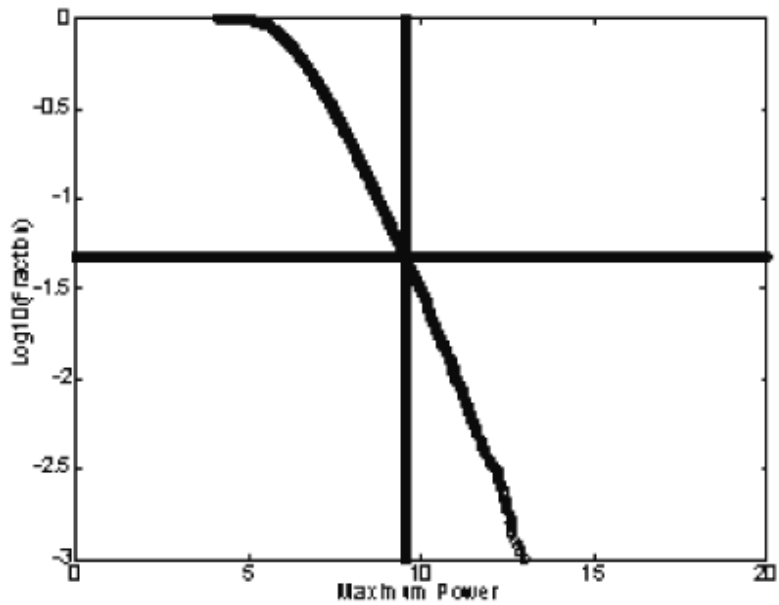

Figure 8. Reverse cumulative distribution function for maximum power, computed by the modified Lomb-Scargle procedure, using the mean live times, over the frequency band 0 to 50 $yr^{-1}$, for 10,000 Monte Carlo simulations of the Super-Kamiokande 5-day data. 477 out of 10,000 simulations have power larger than the actual maximum power (9.56 at frequency 9.43 $yr^1$).

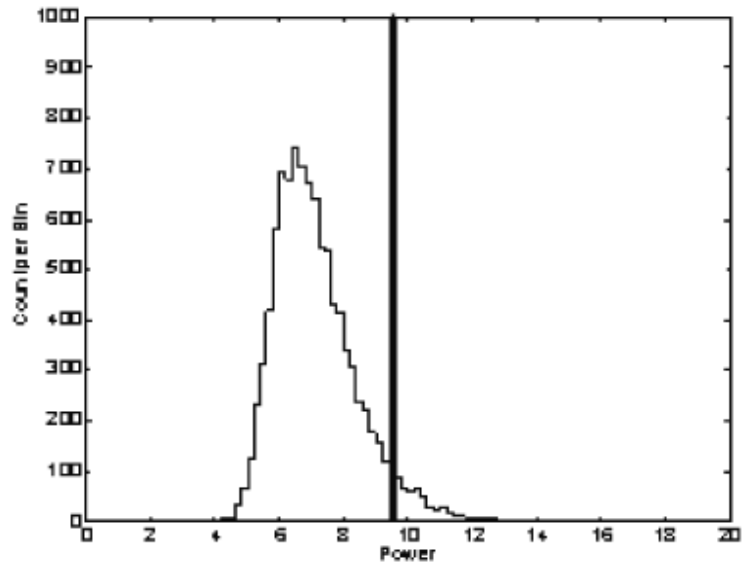

Figure 9. Histogram display of the maximum power, computed by the modified Lomb-Scargle procedure using the mean live times, over the frequency band 0 to 50 yr$^{-1}$, for 10,000 Monte Carlo simulations of the Super-Kamiokande 5-day data. 477 out of 10,000 simulations have power larger than the actual maximum power (9.56 at frequency 9.43 yr$^{-1}$).

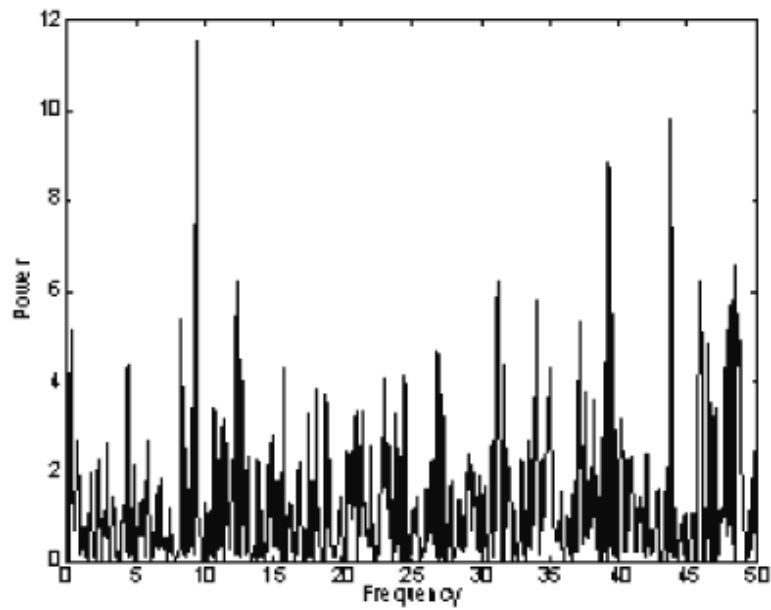

Figure 10. Power spectrum of 5-day Super-Kamiokande data, formed by the SWW likelihood method, using the start and end times.

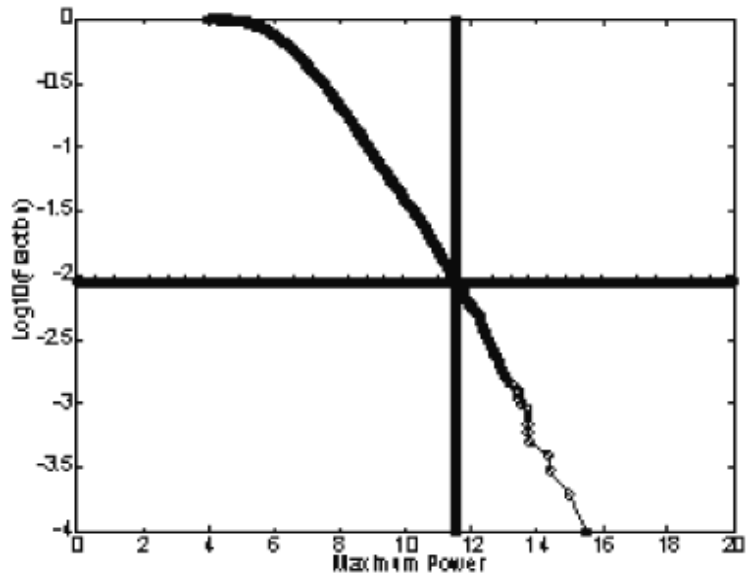

Figure 11. Reverse cumulative distribution function for maximum power formed by the SWW likelihood method, using the start and end times, over the frequency band 0 to 50 yr$^{-1}$, for 10,000 Monte Carlo simulations of the Super-Kamiokande 5-day data. 90 out of 10,000 simulations have power larger than the actual maximum power (11.51 at frequency 9.43 yr$^{-1}$).

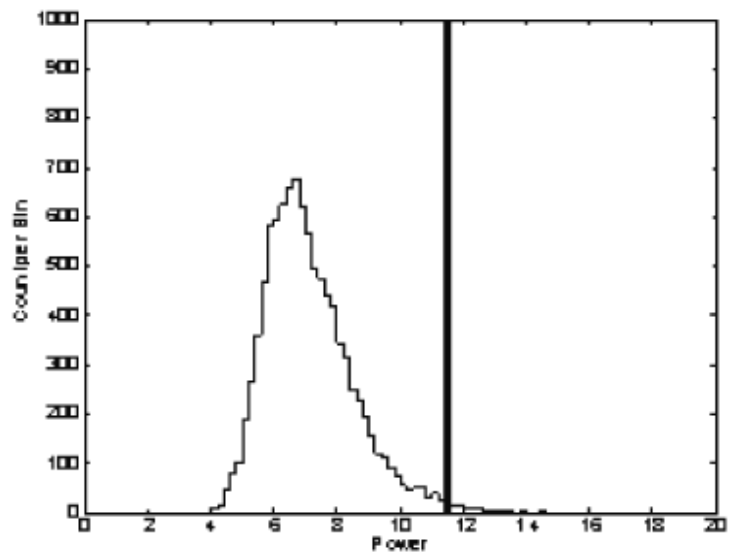

Figure 12. Histogram display of the maximum power, formed by the SWW likelihood method, using the start and end times, over the frequency band 0 to 50 yr$^{-1}$, for 10,000 Monte Carlo simulations of the Super-Kamiokande 5-day data. 90 out of 10,000 simulations have power larger than the actual maximum power (11.51 at frequency 9.43 yr$^{-1}$).

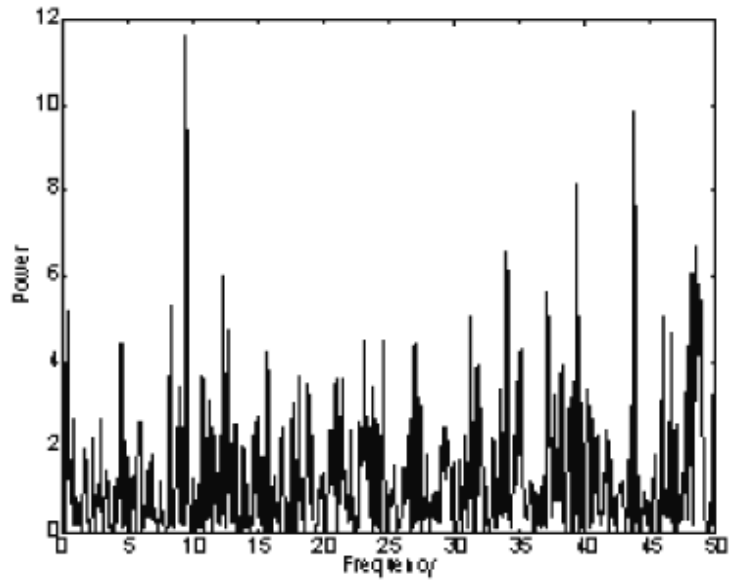

Figure 13. Power spectrum of 5-day Super-Kamiokande data, using the start times, end times, and mean live times, formed by the modified SWW likelihood method.

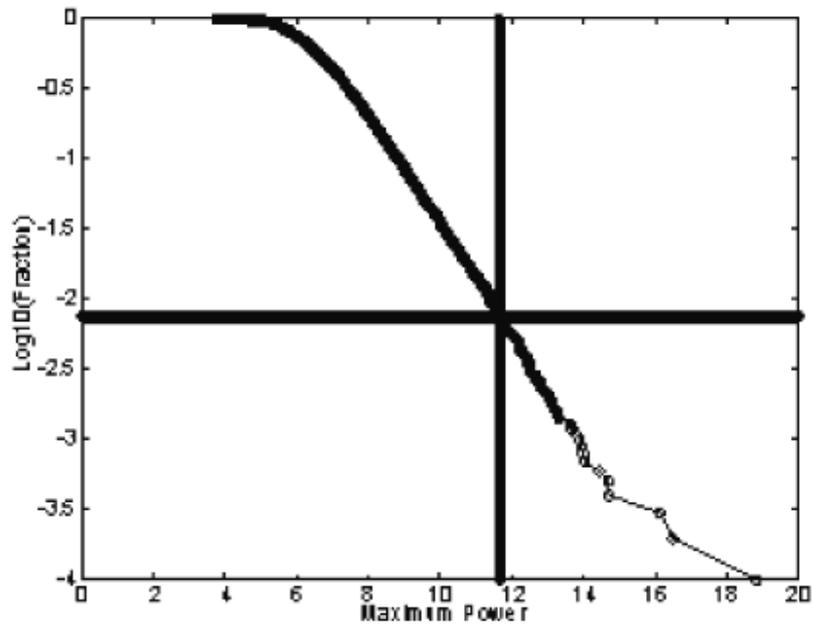

Figure 14. Reverse cumulative distribution function for maximum power, computed by the SWW likelihood method, using the start times, end times, and mean live times, over the frequency band 0 to 50 yr$^{-1}$, for 10,000 Monte Carlo simulations of the Super-Kamiokande 5-day data. 74 out of 10,000 simulations have power larger than the actual maximum power (11.67 at frequency 9.43 yr$^{1}$).

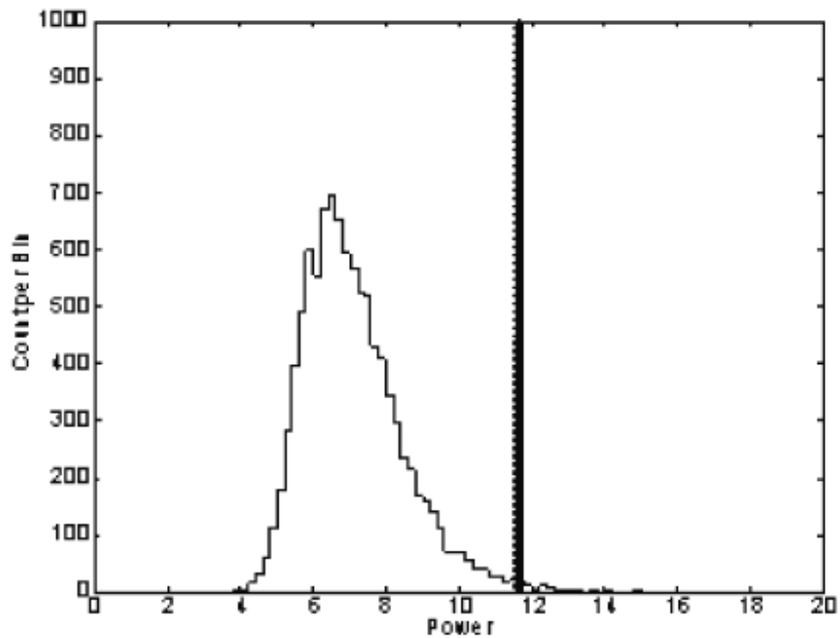

Figure 15. Histogram display of the maximum power, computed by the SWW likelihood method using the start times, end times, and mean live times, over the frequency band 0 to 50 yr$^{-1}$, for 10,000 Monte Carlo simulations of the Super-Kamiokande 5-day data. 74 out of 10,000 simulations have power larger than the actual maximum power (11.67 at frequency 9.43 yr$^{-1}$).

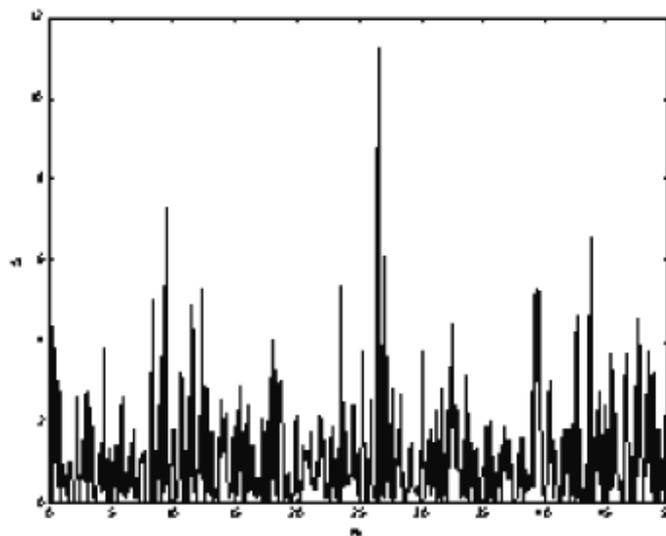

Figure 16. Power spectrum of 10-day Super-Kamiokande data, using the start times, end times, and mean live times, formed by the modified SWW likelihood method.

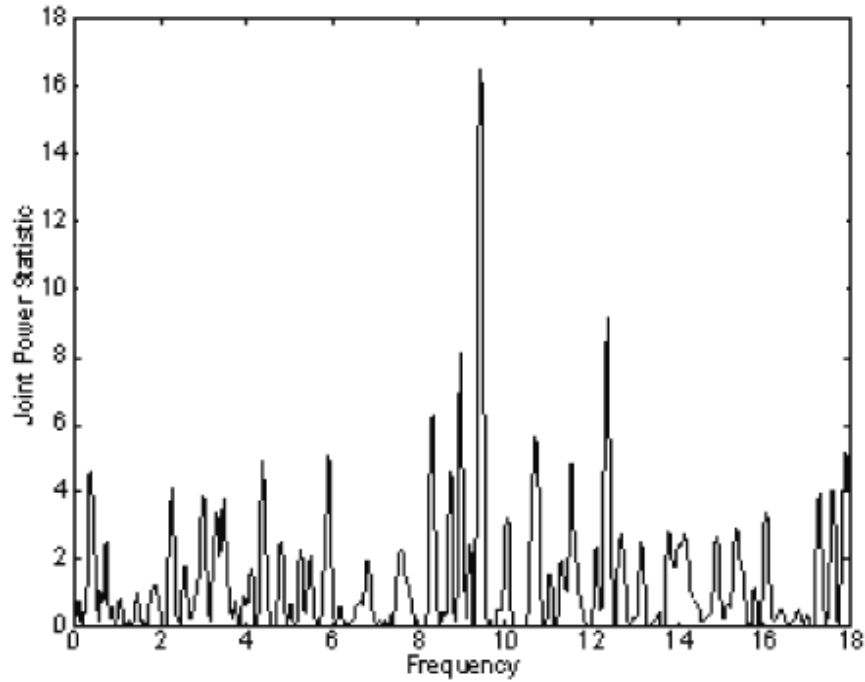

Figure 17. The joint power statistic formed from $S(\nu)$ and $S(\nu_T - \nu)$, over the frequency range 0 to 18 yr$^{-1}$. It is clear that the peaks at 9.42 yr$^{-1}$ and at 26.57 yr$^{-1}$ comprise an alias pair.

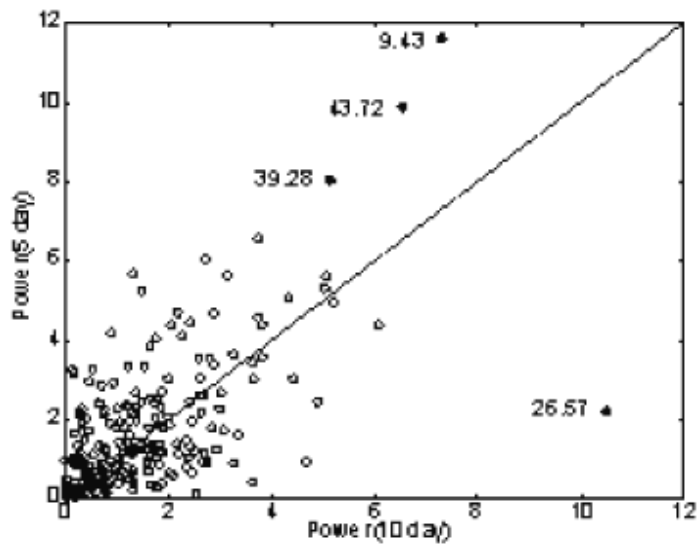

Figure 18. Comparison of five peaks that lie "outside the pack" in a display in which the abscissa values are powers in a spectrum analysis of the 10-day data, and the ordinate values are powers taken from spectrum analysis of the 5-day data. The peaks at 9.42 yr$^{-1}$, 43.72 yr$^{-1}$, and 39.28 yr$^{-1}$ are stronger in the 5-day spectrum than in the 10-day spectrum. The peak at 26.57 yr$^{-1}$ in the 10-day power spectrum is an alias of the peak at 9.42 yr$^{-1}$.

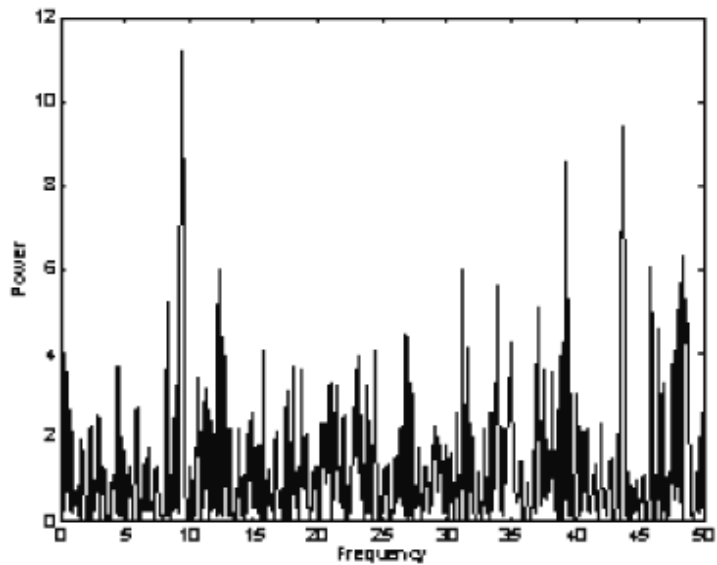

Figure 19. Power spectrum of 5-day Super-Kamiokande data, using the start times and end times, and allowing for a floating offset, formed by the modified SWW likelihood method.

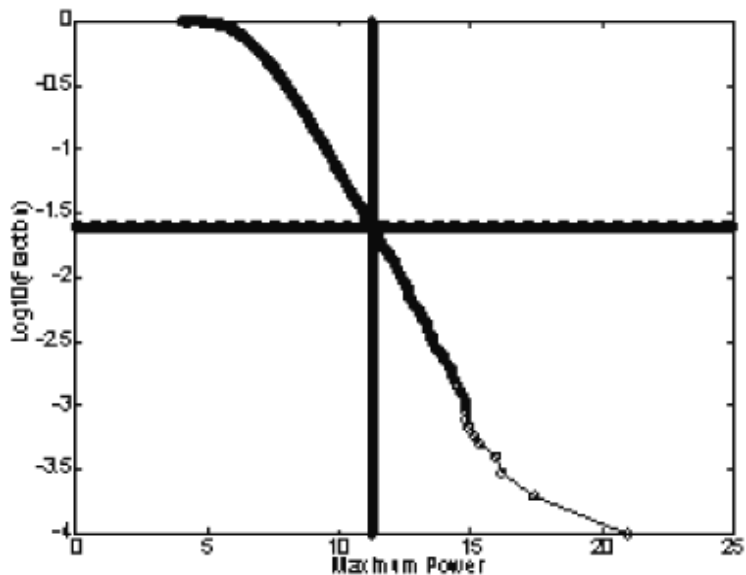

Figure 20. Reverse cumulative distribution function for maximum power, computed by the SWW likelihood method, using the start times and end times and allowing for a floating offset, over the frequency band 0 to 50 $yr^{-1}$, for 10,000 Monte Carlo simulations of the Super-Kamiokande 5-day data. 251 out of 10,000 simulations have power larger than the actual maximum power (11.24 at frequency 9.43 $yr^1$).

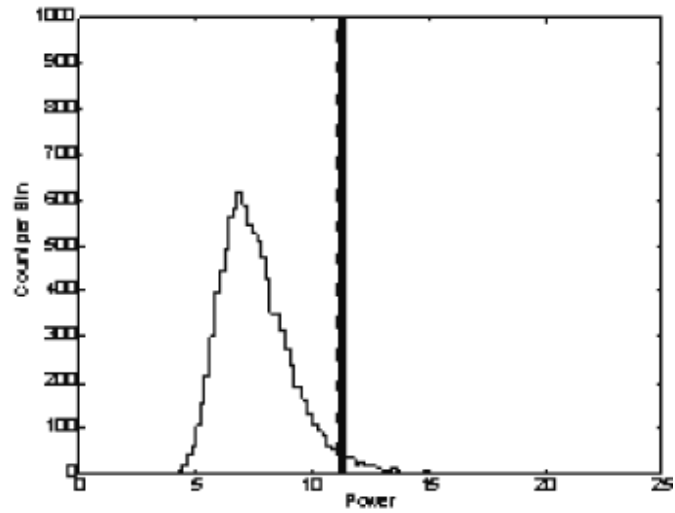

Figure 21. Histogram display of the maximum power, computed by the SWW likelihood method using the start times and end times and allowing for a floating offset, over the frequency band 0 to 50 yr$^{-1}$, for 10,000 Monte Carlo simulations of the Super-Kamiokande 5-day data. 251 out of 10,000 simulations have power larger than the actual maximum power (11.24 at frequency 9.43 yr$^{-1}$).

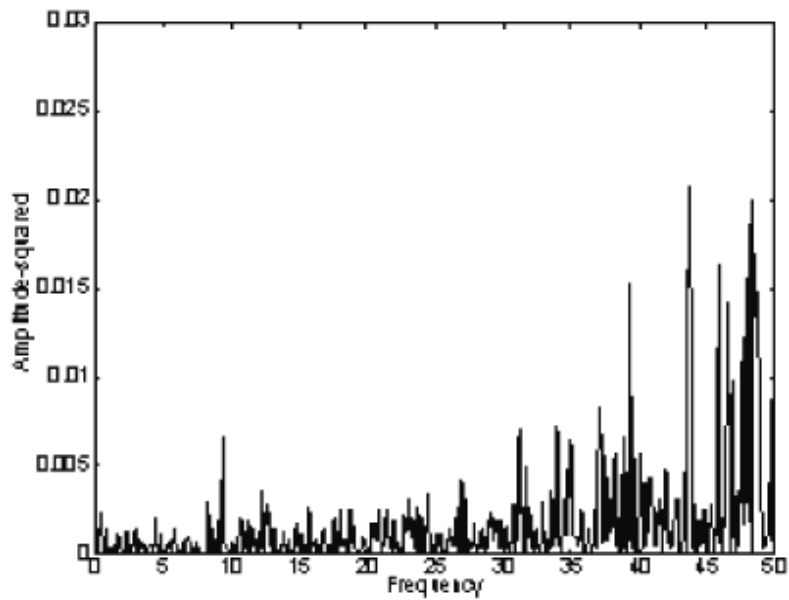

Figure 22. Amplitude squared of the maximum likelihood sinusoidal fit, allowing for a floating offset. There is a large spike at zero frequency (not shown). This and the increase in amplitude with frequency arise from the fact that the equations are ill-conditioned.

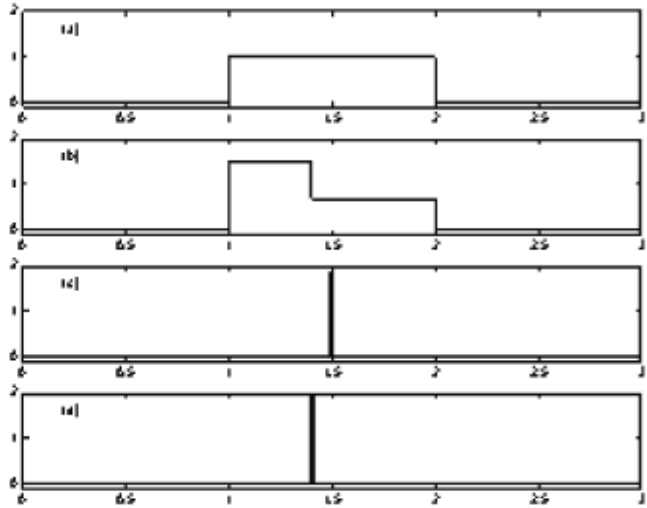

Figure 23. (a) Time window function for uniform weight over start time to end time, as in Section 5. (b) Time window function with non-uniform weight to take account of mean live time, as in Section 6. (c) Delta-function form for time window function at mid-point of bin, as in section 2. (d) Delta-function form for time window function at mean live time, as in Section 3.

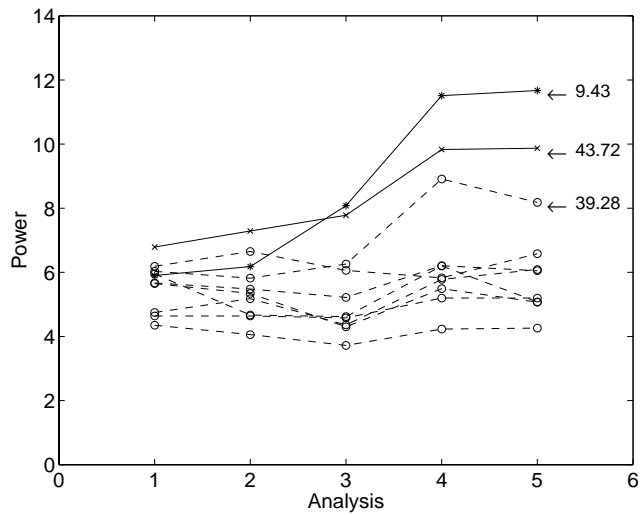

Figure 24. Comparison of powers of top ten peaks for five analysis procedures: (1) basic Lomb-Scargle analysis, mean times; (2) basic Lomb-Scargle analysis, mean live times; (3) modified Lomb-Scargle analysis, mean live times, error data; (4) SWW likelihood method, start times, end times, and error data; (5) SWW likelihood method, start times, end times, mean live times, and error data. Only the peaks at 9.43 $yr^{-1}$ and 43.72 $yr^{-1}$ show a monotonic increase in power.

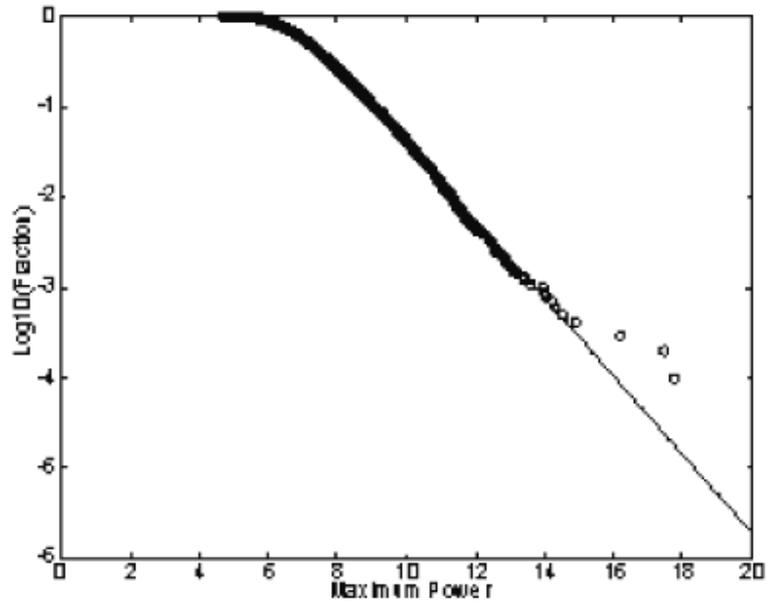

Figure 25. Reverse cumulative distribution function for maximum power, computed by the Lomb-Scargle procedure, using the mean live times, over the frequency band 0 to 70 yr$^{-1}$, for 10,000 Monte Carlo simulations of the Super-Kamiokande 5-day data. We obtain a least-squares fit to the false-alarm formula (A.1) for M = 918. This may be compared with the value 716 assumed by Yoo et al.